\documentclass[aps,prd,preprint,superscriptaddress,amssymb,nofootinbib,showpacs,preprintnumbers]{revtex4}

\usepackage{amsmath}
\usepackage{graphics}
\usepackage{graphicx}
\usepackage[mathcal]{eucal}

\usepackage{epsfig}
\usepackage{dcolumn}
\usepackage{bm}

\newcommand{\eqnref}[1]{Eq.~(\ref{#1})}

\begin{document}

\preprint{COLO-HEP-550, UCI-TR-2009-12}

\title{A Viable Randall-Sundrum Model for Quarks and Leptons with
  $T^{\prime}$ Family Symmetry}

\author{Mu-Chun Chen}
\affiliation{Department of Physics and Astronomy,
University of California, Irvine, CA 92697-4575, USA}

\author{K.T. Mahanthappa}
\affiliation{Department of Physics, University of Colorado 
at Boulder, Boulder, CO 80309-0390, USA}

\author{Felix Yu}
\affiliation{Department of Physics and Astronomy,
University of California, Irvine, CA 92697-4575, USA}

\date{July 22, 2009}


\begin{abstract}
We propose a Randall-Sundrum model with a bulk family symmetry based
on the double tetrahedral group, $T^{\prime}$, which generates the
tri-bimaximal neutrino mixing pattern and a realistic CKM matrix,
including CP violation.  Unlike 4D models where the generation of mass
hierarchy requires additional symmetry, the warped geometry naturally
gives rise to the fermion mass hierarchy through wavefunction
localization.  The $T^{\prime}$ symmetry forbids tree-level
flavor-changing-neutral-currents in both the quark and lepton sectors,
as different generations of fermions are unified into multiplets of
$T^{\prime}$. This results in a low first KK mass scale and thus the
model can be tested at collider experiments.

\end{abstract}





\maketitle


\section{Introduction}
\label{sec:Intro}

The Randall-Sundrum (RS) Model~\cite{Randall:1999ee}, based on a
non-factorizable geometry in a slice of anti-de Sitter ($AdS_5$) space
with a warped background metric, has been proposed as a
non-supersymmetry alternative solution to the gauge hierarchy
problem. In addition to solving the gauge hierarchy problem, the model
can accommodate the fermion mass hierarchy, when the Standard Model
fermions and gauge bosons are allowed to propagate in the
bulk~\cite{bulkfields}.  By localizing different fermions at different
points in the fifth dimension, the widely dispersed masses of the SM
fermions can be accommodated with all 5D Yukawa coupling constants
being order unity~\cite{bulkfermion1,bulkfermion2}. This in turn also
leads to new ways to generate neutrino
masses~\cite{bulkfermion1,Chen:2005mz}. Having the SM particles in the
bulk generally causes large contributions to the electroweak
observables, unless the Kaluza-Klein (KK) mass scale is much higher
than a TeV. To suppress these contributions, realistic models based on
bulk custodial symmetry~\cite{Agashe:2003zs} or large brane kinetic
terms~\cite{branekinetic} have also been built, in which the first KK
mass scale $\sim$ 3 TeV is allowed by the electroweak precision data.

The presence of the 5D bulk mass parameters, which govern the
localizations of the bulk fields, leads to flavor violations in
addition to the contributions caused by the 5D Yukawa
interactions. These two generically independent flavor violation
sources can generate dangerously large
flavor-changing-neutral-currents (FCNCs) already at the tree level
through the exchange of the KK gauge bosons. Although these processes
are suppressed by the built-in RS GIM
mechanism~\cite{bulkfermion2,rsgim1,Cacciapaglia:2007fw}, constraints
from the CP-violating parameter $\epsilon_K$ for
$K^{0}-\overline{K^{0}}$ mixing in the quark sector still give a
stringent bound on the first KK mass scale of $\mathcal{O} (10 \text{
TeV})$~\cite{Bona:2007vi}, when a generic flavor structure is
assumed. Lepton flavor violation (LFV) in various rare leptonic
processes mediated by neutral KK gauge bosons also gives stringent
constraints on the KK mass
scale~\cite{Kitano:2000wr,lfv,Agashe:2006iy}. Even in the absence of
neutrino masses, severe bounds on the first KK mass scale already
arise from processes mediated by tree-level FCNCs, with generic
anarchical 5D Yukawa couplings~\cite{Agashe:2006iy}.

One way to avoid the tree level FCNCs is by imposing minimal flavor
violation (MFV)~\cite{mfv} which assumes that all flavor violation
comes from the Yukawa sector.  Implementation of MFV in the quark
sector has been proposed~\cite{Fitzpatrick:2007sa}. Realizations in
the lepton sector with~\cite{Chen:2008qg} and
without~\cite{Perez:2008ee} a bulk lepton symmetry have also been
suggested. In these implementations, the bulk mass matrices are
properly aligned with the 5D Yukawa matrices as dictated by the
$[U(3)]^{6}$ flavor symmetry. With such alignment, which can arise
from a shining mechanism~\cite{Csaki:2009wc}, tree-level FCNCs can be
suppressed and a first KK mass scale of 2-3 TeV can be allowed,
rendering the model testable at collider experiments~\cite{collider}.
It has also been shown that a low first KK mass scale 
can also be obtained with the so-called minimal flavor protection mechanism~\cite{Santiago:2008vq} which utilized an $U(3)$ flavor symmetry, with non-minimal representations for leptons under $SU(2)_{R}$~\cite{Agashe:2009tu} , or alternatively, by considering a bulk
Higgs and modified value of the 5D strong coupling
constant~\cite{Agashe:2008uz}.

In this paper, we propose an alternative by imposing a bulk family
symmetry. In~\cite{Csaki:2008qq}, a bulk family symmetry based on
$A_4$ has been utilized in the lepton sector. Due to the common bulk
mass term for the three lepton doublets, which is required to generate
tri-bimaximal (TBM) neutrino mixing~\cite{Harrison:1999cf} as
suggested by the recent global fit~\cite{Maltoni:2004ei}, tree-level
leptonic FCNCs are absent.  While $A_4$ well describes the lepton
sector, it does not gives rise to a realistic quark sector. Here we
consider the double tetrahedral
group~\cite{Frampton:1994rk,Feruglio:2007uu,Chen:2007afa,Chen:2009gf},
$T^{\prime}$, as the bulk family symmetry. In addition to
simultaneously giving rise to TBM neutrino mixing and a realistic
Cabibbo-Kobayashi-Maskawa (CKM) matrix, the complex Clebsch-Gordan
(CG) coefficients of $T^{\prime}$ also give the possibility that CP
violation is entirely geometrical in
origin~\cite{Chen:2007afa,Chen:2009gf}. While the three lepton
doublets form a $T^{\prime}$ triplet, as in the case of $A_4$, the
three generations of quarks transform as $2 \oplus 1$, leading to
realistic masses and mixing angles in the quark sector. This
assignment also forbids tree-level FCNCs involving the first and
second generations of quarks, which are the most severely constrained.

The paper is organized as follows. In Sec.~\ref{sec:FV} we review
various sources of flavor violation in generic RS models. We then
present in Sec.~\ref{sec:Model} a RS model with a bulk $T^{\prime}$
family symmetry, in which the tree-level FCNCs are avoided. This is
followed by Sec.~\ref{sec:Results} where our numerical results are
summarized. Sec.~\ref{sec:Conc} concludes the paper.

\section{Flavor Violation in RS}
\label{sec:FV}

In this section, we provide a brief review of flavor violation in
generic Randall-Sundrum models.  We adopt the RS1 framework, where the
fifth dimension $y$ is compactified on a $S^1 / \mathbb{Z}_2$
orbifold.  The resulting bulk geometry between the two orbifold fixed
points corresponds to a slice of $AdS_5$ space of length $\pi R$.  A
3-brane is located at each orbifold fixed point; the geometric warp
factor separating the two branes effects two distinct scales,
$M_{\text{Pl}}$ and $M_{\text{Pl}} e^{-\pi k R}$, where $k \sim
\mathcal{O} (M_{\text{Pl}})$ is the $AdS_5$ curvature scale.  The
electroweak scale naturally arises through the warp factor for $kR
\sim 11$.  Hence, the fundamental scale of the 3-brane located at $y =
0$ is on the order of $M_{\text{Pl}}$, while the fundamental scale of
the 3-brane located at $y = \pi R$ is $\sim M_{\text{Pl}} e^{-\pi k
R}$, which is $\sim \mathcal{O}$(1 TeV).  We confine the Higgs to the
TeV brane and allow the SM fermions and gauge fields to propagate in
the bulk.  In this way, the observable fermion masses and mixings are
determined by the respective wavefunction overlaps between the SM
Higgs and other SM fields on the TeV brane. Here we implicitly assume
the Goldberger-Wise mechanism~\cite{Goldberger:1999uk} for stabilizing
the extra dimension.  

With SM fermions and gauge fields propagating in the bulk, the
electroweak precision measurements place stringent constraints on the
bulk masses of the SM fermions. To satisfy these constraints, among
which the most stringent are the $\rho$ parameter and the $Z$
couplings of the fermions, the bulk mass parameters of the SM fermion
are generally required to be greater than $0.5$.  To preserve the bulk
custodial symmetry, we assume that the bulk obeys $SU(2)_L \times
SU(2)_R \times U(1)_{X}$ symmetry~\cite{Agashe:2003zs}. In addition, to avoid large corrections
to the $Z$ couplings of the fermions, an additional $L
\leftrightarrow R$ parity is required and the fermions must transform in
the non-minimal representations under $SU(2)_{L} \times
SU(2)_{R}$~\cite{Agashe:2006at}. With this assignment, the
$Zb_{L}\overline{b}_{L}$ coupling is protected by the left-right parity 
and consequently the associated bulk parameter can be allowed to be less than $0.5$. While this leads
to a shift in the $Z$ coupling of $t_{L}$, such a deviation is allowed
since experimentally the $Zt_{L}\overline{t}_{L}$ is not very
constrained. For the lighter generations, we take all bulk parameters
to be greater than $0.5$ in our numerical analysis.

The wavefunction overlap depends on the bulk mass parameters
$c_{L_{i}}$ and $c_{R_{j}}$ according to the function
\begin{equation}
f(c_{L_{i}}, c_{R_{j}}) = \frac{1}{2} 
\sqrt{ \frac{ (1 - 2 c_{L_{i}}) (1 - 2 c_{R_{j}})}{ (e^{(1 - 2
c_{L_{i}}) \pi k R} - 1)(e^{(1 - 2 c_{R_{j}}) \pi k R} - 1)} } 
e^{(1 - c_{L_{i}} - c_{R_{j}}) \pi k R} \ ,
\label{eqn:ffactor}
\end{equation}
where the first factors are from normalization, and the extra $e^{\pi
  kR}$ is from the canonical normalization of the Higgs kinetic term.
For $c_{L_{i}},\ c_{R_{j}} > 0.5$, the fermion fields are localizated
toward the Planck brane and have small wavefunction overlaps with the
Higgs field at the TeV brane.  On the other hand, for $c_{L_{i}},
\ c_{R_{j}} < 0.5$, the fields will have large wavefunction overlaps
with the Higgs.  This wavefunction localization
mechanism~\cite{bulkfermion1,bulkfermion2,Chen:2005mz} can naturally
give rise to the observed fermion mass hierarchies.

Even though this is a natural way to generate the mass hierarchy, the
non-universal bulk mass terms for the three generations of fermions
generally lead to tree-level FCNCs.  Consider the 4D effective gauge
coupling for fermions from the kinetic term after integrating out the
fifth coordinate $y$,
\begin{equation}
\mathcal{L}^{\text{4D}}_{\text{Kin}} \supset \int dy e^{-4k \left| y \right|} 
i \overline{\Psi} \gamma^M D_M \Psi \rightarrow g G 
\overline{\psi}
\left( \begin{array}{ccc}
f(c_1, c_1)^2 & 0 & 0 \\
0 & f(c_2, c_2)^2 & 0 \\
0 & 0 & f(c_3, c_3)^2 \\
\end{array} \right) \psi
\end{equation}
where $M = \{\mu, 5 \}$ labels the coordinates with $\mu$ as the usual
4D Lorentz index.  Schematically, $\Psi$ denotes the 5D fermion field,
$\psi = \left( \begin{array}{ccc} \psi_1 & \psi_2 & \psi_3
\end{array} \right)^T$ is the three-generation 4D fermion field
in the gauge basis, $g$ is the gauge coupling, and $G$ is the gauge
field in the adjoint representation.  When the above gauge
interactions are rotated to the mass basis by insertion of
$V^{\dagger} V$, where the mass eigenstates are $\psi_m = V \psi$, the
4D effective gauge interactions become
\begin{eqnarray}
\lefteqn{
g G \overline{\psi} V^{\dagger} V \left( \begin{array}{ccc}
f(c_1, c_1)^2 & 0 & 0 \\
0 & f(c_2, c_2)^2 & 0 \\
0 & 0 & f(c_3, c_3)^2 \\
\end{array} \right) V^{\dagger} V \psi } \nonumber \\
&=& g G \overline{\psi}_m V \left( \begin{array}{ccc}
f(c_1, c_1)^2 & 0 & 0 \\
0 & f(c_2, c_2)^2 & 0 \\
0 & 0 & f(c_3, c_3)^2 \\
\end{array} \right) V^{\dagger} \psi_m \nonumber \\
&\equiv& g G \overline{\psi}_m M \psi_m \ .
\label{eqn:MFCNC}
\end{eqnarray}
Thus, if any two bulk mass terms are unequal, then non-zero
off-diagonal elements in the gauge coupling matrix, $M$, in the mass
basis can generally be present and hence sizable FCNC transitions may
be generated.

There are two distinct ways to alleviate this problem.  First, if
there is flavor universality, which our model possesses for the
left-handed lepton doublets in order to generate TBM neutrino mixing,
then $c_1 = c_2 = c_3$ and the $V$ and $V^{\dagger}$ unitary matrices
commute through~\eqnref{eqn:MFCNC}, leaving $M \propto 1_{3\times 3}$
without neutral flavor-changing transitions.  On the other hand, if
the gauge basis can be freely rotated to coincide with the mass basis
(alignment), as will be true for the right-handed leptons in our
model, then $V = 1_{3 \times 3}$ and again, $M$ will contain no
off-diagonal entries.  We note that a combination of these two ideas
is also useful.  The three generations of quarks in our model, for
example, transform as the $2 \oplus 1$ representations under the
double tetrahedral group, $T^{\prime}$, as in the usual 4D
models~\cite{Frampton:1994rk,Feruglio:2007uu,Chen:2007afa,Chen:2009gf}. This
representation assignment is required in order to generate realistic
quark masses and mixing pattern. The universality that is exhibited
among the first two generations forbids the tree-level flavor
transitions that cause $K^{0}-\overline{K^{0}}$ mixing.  The
near-identity CKM matrix mimics alignment, and so the resulting
neutral current matrix possesses minimal 1--3, 2--3, 3--1, and 3--2
couplings, which are still allowed by current experimental
constraints.  Our model's features of flavor universality for
left-handed (LH) leptons, alignment for right-handed (RH) leptons, and
the $2 \oplus 1$ framework for quarks all result from the $T^{\prime}$
bulk family symmetry.  We summarize the relevant properties of the
finite group $T^{\prime}$ in Appendix~\ref{app:Tp}. We comment that
since the family group $T^{\prime}$ is a direct product group with the
enlarged $SU(2)_{L} \times SU(2)_{R}$ symmetry, assigning the SM
fermions in the non-minimal representations under the LR group as
required by avoiding the EW precision constraints does not affect our
analysis.

We note that the universality in the $Z$ couplings to the fermions may
be spoiled by higher order effects. The leading higher order effects
are dim-6 operators which can in general be induced by (i) the mixing
between the fermion zero mode and its KK modes; (ii) the mixing
between the zero mode and KK modes of the $Z$ boson. In addition,
there can be loop contributions to flavor violating $Z$ couplings in
the presence of brane localized kinetic terms~\cite{Santiago:2008vq}. 
These non-universal kinetic terms are induced by the brane-localized 
Yukawa couplings, and they could lead to loop-suppressed non-universal 
shifts in the normalization of the zero mode wave functions. 
Since these loop contributions correspond to dim-8 operators, these
effects are sub-dominant in the presence of non-vanishing
contributions from dim-6 operators. In our model, due to the
$T^{\prime}$ symmetry which leads to universal bulk mass parameters
for the lighter quarks, the fermion couplings to the
$Z$ boson induced by higher order effects due to the mixing of the $Z$
boson zero mode and its KK modes are flavor-preserving.  As a result
the leading contributions to flavor violating $Z$ couplings are due to
the dim-6 operators induced by the mixing of fermion zero mode and its
KK modes.  An estimate of such higher order effects in our model is
presented in Section \ref{sec:Results}. Our estimate shows that with a
low first KK mass scale of 3-4 TeV, these higher order contributions
are suppressed enough to satisfy all experimental
constraints~\cite{Santiago:2008vq,Csaki:2009wc}.

\section{The Model}
\label{sec:Model}

In our model, we impose the discrete $T^{\prime}$ symmetry as a flavor
symmetry for SM fermion fields placed in the bulk.  As mentioned
previously, the SM Higgs is confined to the TeV brane, while the SM
gauge and fermion fields are allowed to propagate in the bulk. The
three generations of LH lepton doublets $L$ and three generations of
RH neutrinos $N$ are unified into triplet representations, and the RH
charged leptons $e$, $\mu$, $\tau$ transform as inequivalent
one-dimensional representations. The first two generations of LH
quarks $Q_{12}$, RH up-type quarks $U$, and RH down-type quarks $D$
are each in $T'$ doublet representations, while the third generation
LH quark doublet $Q_3$, RH top quark $T$, and RH bottom quark $B$, and
the SM Higgs field, $H$, are all pure singlets under $T^{\prime}$. To
break the $T'$ symmetry, we need a set of flavons, which are all
singlets under the SM gauge group.  The representation assignments for
the SM fermions and $T'$ flavons are summarized in Table
\ref{table:Tpassign}.

\begin{table}[tb]
\begin{tabular}{|c| |c|c|c|c|c| |c|c|c| |c|c|c|c|c|c| 
|c|c|c|c|c|c|c|c| }
\hline
   & $L$ & $N$ & $e$ & $\mu$ & $\tau$ 
    & $\phi$ & $\phi^{\prime}$ & $\sigma$ & $Q_{12}$ & $Q_3$ 
    & $U$ & $T$ & $D$ & $B$ & $\alpha$ & $\beta$ & $\zeta$ & $\xi$ 
    & $\chi_U$ & $\chi_D$ & $\eta_U$ & $\eta_D$ \\
\hline $T^{\prime}$ 
   & 3   & 3   & 1   & 1$^{\prime\prime}$ & 1$^{\prime}$ 
    & 3      & 3               & 1        & 2       & 1     
    & 2   & 1   & 2   & 1   & 3        & 3       & 1       & 1     
    & 2        & 2        & 2        & 2        \\
\hline
\end{tabular}
\caption{Representation assignments for SM fermion and $T^{\prime}$
flavon fields.  The field definitions are given the main text.}
\label{table:Tpassign}
\end{table}

\subsection{The Lepton Sector}

The 5D Lagrangian involving leptons, including the canonical 5D
kinetic term, $\mathcal{L}^{\text{lep}}_{\text{Kin}}$, the 5D bulk
mass term, $\mathcal{L}^{\text{lep}}_{\text{Bulk}}$, and the 5D Yukawa
interactions for charged leptons,
$\mathcal{L}^{\text{lep}}_{\text{Yuk, } \ell}$ and neutrinos
$\mathcal{L}^{\text{lep}}_{\text{Yuk, } \nu}$, is
\begin{equation}
\mathcal{L}^{\text{lep}}_{\text{5D}} \supset
\mathcal{L}^{\text{lep}}_{\text{Kin}} +
\mathcal{L}^{\text{lep}}_{\text{Bulk}} +
\mathcal{L}^{\text{lep}}_{\text{Yuk, } \ell} +
\mathcal{L}^{\text{lep}}_{\text{Yuk, } \nu} \ .
\label{eqn:Llep}
\end{equation}
The bulk mass terms in our model are given by
\begin{equation}
\mathcal{L}^{\text{lep}}_{\text{Bulk}} = k \left( \overline{L} c_L L +
\overline{e} c_e e + \overline{\mu} c_{\mu} \mu + \overline{\tau}
c_{\tau} \tau + \overline{N} c_N N \right) \ ,
\label{eqn:Llepmass}
\end{equation}
where the bulk mass terms $c_i$ are dimensionless.  Due to the
$T^{\prime}$ symmetry, the number of bulk mass terms is significantly
reduced when compared to the generic case without a family symmetry.

We now distinguish the charged lepton Yukawa interactions from the
neutrino Yukawa interactions because in our model, neutrinos can be
treated in two ways: (i) a purely Dirac mass structure or (ii) a type
I seesaw realization.  For the Dirac (Dc) neutrino case, all 5D Yukawa
interactions take place on the TeV brane, where the SM Higgs and the
flavon fields are confined.  The charged lepton interactions are then
\begin{eqnarray}
& & \mathcal{L}^{\text{lep}}_{\text{Yuk, } \ell} = \delta 
\left( y - \pi R \right) \left[ \frac{1}{k} \overline{H}(x)
\left( y_e^{\text{5D}} \overline{L}(x,y) e(x,y) \frac{\phi(x)}{\Lambda}
\right. \right. \nonumber \\
& & \left. \left.
+y_{\mu}^{\text{5D}} \overline{L}(x,y) \mu(x,y) \frac{\phi(x)}{\Lambda}
+y_{\tau}^{\text{5D}} \overline{L}(x,y) \tau(x,y) 
\frac{\phi(x)}{\Lambda} \right) \right] + \text{ h.c.}
\label{eqn:Llepyuk}
\end{eqnarray}
and the purely Dirac neutrino mass structure is
\begin{eqnarray}
& & \mathcal{L}^{\text{lep}}_{\text{Yuk, } \nu, \text{ Dc}} = \delta 
\left( y - \pi R \right) \nonumber \\
& & \left[ \frac{1}{k} H(x) \overline{L}(x,y) N(x,y) 
\left( y_{\nu, \text{ Dc, a}}^{\text{5D}}
\frac{\phi_{\text{ Dc}}^{\prime}(x)}{\Lambda} 
+ y_{\nu, \text{ Dc, b}}^{\text{5D}} 
\frac{\sigma_{\text{ Dc}} (x)}{\Lambda}
\right) \right] + \text{ h.c.} \ ,
\label{eqn:Lnuyuk1}
\end{eqnarray}
where the 5D Yukawa coupling constants $y_i^{\text{5D}}$ are
dimensionless.  For this Dirac neutrino case, the flavon fields
$\phi$, $\phi_{\text{Dc}}^{\prime}$ and $\sigma_{\text{Dc}}$ are
scalar fields confined to the TeV brane, and therefore the cutoff
scale $\Lambda$ of the higher dimensional operators in the above
equation is $\Lambda \sim \mathcal{O}(1 \text{ TeV})$.

On the other hand, in our seesaw (SS) realization, the heavy Majorana
RH neutrino mass term arises from flavon fields confined to the Planck
brane and a Dirac mass contribution from the SM Higgs confined to the
TeV brane.  While the charged lepton Yukawa interactions are unchanged
from~\eqnref{eqn:Llepyuk}, the neutrino Yukawa terms are now replaced
by
\begin{eqnarray}
& & \mathcal{L}^{\text{lep}}_{\text{Yuk, } \nu, \text{ SS}} = 
\left\{ \delta \left( y \right) \left[ \frac{1}{k} N^T (x,y) N(x,y) 
\left( y_{\nu, \text{ SS, a}}^{\text{5D}} 
\phi_{\text{SS}}^{\prime}(x') 
+ y_{\nu, \text{ SS, b}}^{\text{5D}} \sigma_{\text{SS}}(x') 
\right) \right] \right. \nonumber \\
& & \left. 
+ \delta \left( y - \pi R \right) \left[ \frac{1}{k} H(x) 
y_{\nu, \text{ SS, c}}^{\text{5D}} \overline{L}(x,y) N(x,y) 
\right] \right\} + \text{ h.c.} \ ,
\label{eqn:Lnuyuk2}
\end{eqnarray}
where $x'$ is the 4D spacetime coordinate on the Planck brane, and the
flavon fields $\phi_{\text{SS}}^{\prime}$ and $\sigma_{\text{SS}}$
have the same $T^{\prime}$ representations as their pure Dirac
counterparts in~\eqnref{eqn:Lnuyuk1} and hence we use the same
notation in both equations.  The relevant $T^{\prime}$ breaking scale
for these flavons, however, is $\Lambda_{\text{UV}} \sim 10^{19}$ GeV,
not the TeV scale as before.

We emphasize for both cases that, in order to generate TBM mixing
pattern for the neutrinos, the $L$ and $N$ transform as triplets under
$T^{\prime}$.  (We note that in the Dirac neutrino case, we choose the
coefficient of the contraction of $L$ and $N$ that gives the
antisymmetric triplet $3_A$ to be zero.)  This choice of $T^{\prime}$
representations also ensures that the bulk mass matrices $c_L$ and
$c_N$ each become universal among the three generations, and hence our
model avoids tree-level FCNCs in the lepton sector.

The flavon fields in~\eqnref{eqn:Llepyuk} and~\eqnref{eqn:Lnuyuk1}
acquire vacuum expectation values (VEVs) along the following
directions
\begin{equation}
\langle \phi \rangle = \phi_0 \Lambda \left( \begin{array}{c} 1 \\ 0
\\ 0 \\
\end{array} \right) \qquad 
\langle \phi_{\text{Dc}}^{\prime} \rangle = 
\phi_{0, \text{ Dc}}^{\prime} \Lambda \left( 
\begin{array}{c} 1 \\ 1 \\ 1 \\ \end{array} \right) \qquad
\langle \sigma_{\text{Dc}} \rangle = \sigma_{0, \text{ Dc}} \Lambda \ .
\label{eqn:lepvev1}
\end{equation}
For this Dirac neutrino case, we explicitly factor out the
$T^{\prime}$ breaking scale $\Lambda$ in order to leave the
coefficients, $\phi_0$, $\phi_{0 \text{ Dc}}^{\prime}$, and $\sigma_{0
\text{ Dc}}$, dimensionless.  For the seesaw case, the VEV of the $\phi$ field is
unchanged, but the VEVs of the $\phi^{\prime}$ and $\sigma$ fields become 
\begin{equation}
\langle \phi_{\text{SS}}^{\prime} \rangle = 
\phi_{0, \text{ SS}}^{\prime} \Lambda_{\text{UV}} \left(
\begin{array}{c} 1 \\ 1 \\ 1 \\ \end{array} \right) \qquad
\langle \sigma_{\text{SS}} \rangle = \sigma_{0, \text{ SS}}
\Lambda_{\text{UV}} \ ,
\label{eqn:lepvev2}
\end{equation}
where $\Lambda_{\text{UV}}$ is the relevant $T^{\prime}$ breaking
scale for these flavons.

Upon $T^{\prime}$ symmetry breaking due to $\langle \phi \rangle$ 
in~\eqnref{eqn:lepvev1}, the charged lepton mass matrix becomes
\begin{equation}
M_e = v \phi_0 \left( \begin{array}{ccc}
y_e & 0 & 0 \\
0 & y_\mu & 0 \\
0 & 0 & y_\tau \\
\end{array} \right) \ ,
\label{eqn:Lmatrix}
\end{equation}
where $v = 246$ GeV is the SM Higgs VEV, and the effective 4D Yukawa
coupling constants
\begin{equation}
y_\ell = y_\ell^{\text{5D}} f(c_L, c_\ell) \ ,
\label{eqn:yuk_lep}
\end{equation}
for $\ell = e$, $\mu$, $\tau$, depend on the wavefunction profiles of
the fermions as characterized by the overlap function $f(c_L, c_\ell)$
defined in~\eqnref{eqn:ffactor}. We remark that even with universal
$c_L$ for the lepton doublets, it is clear that the observed charged
lepton mass hierarchy can be obtained via the non-universal values for
the bulk mass parameters for the RH charged leptons, $c_\ell$.

For the case of pure Dirac neutrinos, the Lagrangian
in~\eqnref{eqn:Lnuyuk1} and VEVs in~\eqnref{eqn:lepvev1} lead to the
4D effective neutrino mass matrix
\begin{equation}
M_\nu = f(c_L, c_N) v \left( \begin{array}{ccc}
2A + B & -A & -A \\
-A & 2A & B - A \\
-A & B - A & 2A \\
\end{array} \right) \ ,
\label{eqn:Nmatrix1}
\end{equation}
where $f(c_i, c_j)$ is defined in~\eqnref{eqn:ffactor} and the
parameters $A$ and $B$ are
\begin{equation}
A = \frac{1}{3} y_{\nu, \text{ Dc, a}}^{\text{5D}} 
\phi_{0, \text{ Dc}}^{\prime} \qquad 
B = y_{\nu, \text{ Dc, b}}^{\text{5D}} \sigma_{0, \text{ Dc}} \ .
\end{equation}
This mass matrix is form-diagonalizable by the TBM mixing matrix, 
\begin{equation}
U_{\text{TBM}} = \left( \begin{array}{ccc}
 \sqrt{2/3} & \sqrt{1/3} & 0 \\
-\sqrt{1/6} & \sqrt{1/3} & -\sqrt{1/2} \\
-\sqrt{1/6} & \sqrt{1/3} &  \sqrt{1/2} \\
\end{array} \right) \ ,
\label{eqn:TBM}
\end{equation}
independent of the values of $A$ and $B$. The three neutrino mass
eigenvalues depend on $A$ and $B$, and the wavefunction overlap sets
the overall scale of the neutrino masses. This leads to the following
predictions for the absolute masses of the three neutrinos
\begin{equation}
M_{\nu}^D = f(c_L, c_N) v \text{ diag} (3A + B, B, 3A - B) \ ,
\end{equation}
which obey the sum rule $m_1 - m_3 = 2 m_2$~\cite{Chen:2008eq}.  As a
consequence of the fact that the solar mixing angle $\Delta
m_{\text{sol}}^2 = \Delta m_{21}^2$ is known to be positive, this
model predicts a normal hierarchy ordering~\cite{Chen:2008eq}.
Correspondingly, the expressions for neutrino mass squared differences
are
\begin{equation}
\Delta m_{\text{sol}}^2 = \Delta m_{21, \text{ Dc}}^2 
= - \left( f(c_L, c_N) v \right)^2 (9A^2 + 6 AB)
\end{equation}
and
\begin{equation}
\Delta m_{\text{atm}}^2 = \left| \Delta m_{31, \text{ Dc}}^2 \right| 
= \left| -\left( f(c_L, c_N) v \right)^2 12 AB \right| \ .
\end{equation}

For our seesaw realization, the neutrino sector has the block mass
matrix form
\begin{equation}
M_{\nu} = \left( \begin{array}{cc}
0 & M_{\text{Dc}}^T \\
M_{\text{Dc}} & M_{\text{RR}} \\
\end{array} \right) \ ,
\label{eqn:Mnublock}
\end{equation}
where each entry is understood to be a $3 \times 3$ matrix.  With the
Lagrangian in~\eqnref{eqn:Lnuyuk2} and VEVs in~\eqnref{eqn:lepvev2},
the 4D effective matrices in~\eqnref{eqn:Mnublock} are
\begin{equation}
M_{\text{RR}} = \frac{1 - 2 c_N}{2 \left( e^{(1 - 2 c_N) \pi k R} 
- 1 \right)} \Lambda_{\text{UV}}
\left( \begin{array}{ccc}
2A + B & -A & -A \\
-A & 2A & B - A \\
-A & B - A & 2A \\
\end{array} \right) \ ,
\label{eqn:Nmatrix2a}
\end{equation}
where 
\begin{equation}
A = \frac{1}{3} y_{\nu, \text{ SS, a}}^{\text{5D}} 
\phi_{0, \text{ SS}}^{\prime} \qquad
B = y_{\nu, \text{ SS, b}}^{\text{5D}} \sigma_{0, \text{ SS}} \ ,
\end{equation}
and
\begin{equation}
M_{\text{Dc}} = y_{\nu, {\text{ SS, c}}}^{\text{5D}} f(c_L, c_N) v
\left( \begin{array}{ccc}
1 & 0 & 0 \\
0 & 0 & 1 \\
0 & 1 & 0 \\
\end{array} \right) \ .
\label{eqn:Nmatrix2b}
\end{equation}

Given this seesaw structure, the resulting light effective LH neutrino
mass matrix is thus~\cite{Minkowski:1977sc}
\begin{equation}
M_{\nu, \text{ eff}} = M_{\text{Dc}} M_{\text{RR}}^{-1}
M_{\text{Dc}}^T
\end{equation}
which is diagonalized by the TBM mixing matrix~\cite{Chen:2009um} to
give eigenvalues
\begin{equation}
M_{\nu, \text{ eff}}^D = 
\frac{ \left( y_{\nu, \text{ SS, c}}^{\text{5D}}
 v \right)^2 }{2 \Lambda_{\text{UV}} } 
\frac{ 1 - 2 c_L}{e^{(1 - 2 c_L) \pi k R} - 1}
e^{2 (1 - c_L - c_N) \pi k R} \text{ diag} \left( \frac{1}{3A + B},
\frac{1}{B}, \frac{1}{3A - B} \right) \ .
\end{equation}
Unlike the Dirac neutrino case where the mass eigenvalues predicted a
normal hierarchy, the seesaw realization can accomodate either a
normal or inverted hierarchy.  For either hierarchy, we have
\begin{equation}
\Delta m_{\text{sol}}^2 = \Delta m_{21, \text{ SS}}^2 = 
\kappa^2 \left[ \frac{1}{A^2} - \frac{1}{\left( 3A + B \right)^2}
\right] 
\end{equation}
and
\begin{equation}
\Delta m_{\text{atm}}^2 = \left| \Delta m_{31, \text{ SS}}^2 \right| = 
\left| \kappa^2 \left[ \frac{1}{\left( 3A - B \right)^2} - 
\frac{1}{\left( 3A + B \right)^2} \right] \right| \ ,
\end{equation}
with 
\begin{equation}
\kappa = \frac{ \left( y_{\nu, \text{ SS, c}}^{\text{5D}} v
\right)^2 }{2 \Lambda_{\text{UV}} } \frac{ 1 - 2 c_L}{e^{(1 - 2 c_L)
\pi k R} - 1} e^{2 (1 - c_L - c_N) \pi k R} \ .
\end{equation}

\subsection{The Quark Sector}

In the quark sector, since the top quark is heavy, it suggests that
its mass is allowed by the $T^{\prime}$ symmetry and thus can be
generated at the renormalizable level.  The lighter generations, on
the other hand, have mass terms which are generated after the breaking
of the $T^{\prime}$ symmetry. Their mass terms hence are generated by
higher-dimensional operators and are suppressed by the (IR)
$T^{\prime}$ breaking scale, $\Lambda$.  These considerations
therefore suggest the $2 \oplus 1$ representation assignment, whereby
the first two generations of quarks form a doublet of $T^{\prime}$ and
the third generation transforms as a pure singlet.  The 5D Lagrangian
involving quarks is given by
\begin{equation}
\mathcal{L}^{\text{qrk}}_{\text{5D}} \supset
\mathcal{L}^{\text{qrk}}_{\text{Kin}} +
\mathcal{L}^{\text{qrk}}_{\text{Bulk}} +
\mathcal{L}^{\text{qrk}}_{\text{Yuk}} \ ,
\label{eqn:Lqrk}
\end{equation}
where $\mathcal{L}^{\text{qrk}}_{\text{Kin}}$ is the canonical
5D kinetic term and the 5D bulk mass terms are 
\begin{equation}
\mathcal{L}^{\text{qrk}}_{\text{Bulk}} = k \left( \overline{Q}_{12}
c_{Q_{12}} Q_{12} + \overline{Q}_3 c_{Q_3} Q_3 + \overline{U} c_U U +
\overline{T} c_T T + \overline{D} c_D D + \overline{B} c_B B \right)
\ .
\label{eqn:Lqrkmass}
\end{equation}
Similar to the lepton sector, due to the $T^{\prime}$ family symmetry
and the $2 \oplus 1$ structure, the number of bulk parameters is
greatly reduced in our model.

Without further symmetries or assumptions, the most general Yukawa
interactions within the $2 \oplus 1$ framework has at most eight
flavon fields, which we will now demonstrate.  The most general
$T^{\prime}$ invariant quark Yukawa terms can be written as
\begin{eqnarray}
\mathcal{L}^{\text{qrk}}_{\text{Yuk}} &=& 
\delta \left( y - \pi R \right) \left\{ \frac{1}{k} H(x) \left[
  y_{12}^U \overline{Q}_{12}(x,y) U(x,y) 
          \left(\frac{\alpha(x) + \zeta(x)}{\Lambda} \right) +
  y_3^U \overline{Q}_3(x,y) U(x,y) \frac{\chi_U(x)}{\Lambda} 
\right. \right. \nonumber \\
& & \left. + 
  y_{12}^T \overline{Q}_{12}(x,y) T(x,y) \frac{\eta_U(x)}{\Lambda} +
  y_3^T \overline{Q}_3(x,y) T(x,y) \right] \nonumber \\
& & + \frac{1}{k} \overline{H}(x) \left[
  y_{12}^D \overline{Q}_{12}(x,y) D(x,y) 
          \left(\frac{\beta(x) + \xi(x)}{\Lambda} \right) +
  y_3^D \overline{Q}_3(x,y) D(x,y) \frac{\chi_D(x)}{\Lambda} 
\right. \nonumber \\ 
& & \left. \left. + 
  y_{12}^B \overline{Q}_{12}(x,y) B(x,y) \frac{\eta_D(x)}{\Lambda} +
  y_3^B \overline{Q}_3(x,y) B(x,y) \right] \right\} \ .
\label{eqn:Lqrkyuk}
\end{eqnarray}
In principle, there are two separate Yukawa couplings for the triplet
and singlet flavon fields, $\alpha$ and $\zeta$ (and similarly for
$\beta$ and $\xi$). Nevertheless, the difference between the two
couplings can be absorbed into the values of the VEVs of $\alpha$ and
$\zeta$. With this re-scaling freedom, we can assume the same coupling
constant for both terms.  The flavon fields $\alpha$, $\zeta$,
$\chi_U$, $\eta_U$ (as well as their down-type counterparts $\beta$,
$\xi$, $\chi_D$, $\eta_D$) transform under $T^{\prime}$ as $3$, $1$,
$2$, and $2$, respectively, as reflected in Table
\ref{table:Tpassign}.  Since every possible SM field contraction is
exercised, introducing new flavon fields in the same representations
as above would be redundant.

We continue our derivation of the most general $2 \oplus 1$ quark mass
matrix and present the up-type quarks, seeing that the down-type
matrix will be exactly analogous.  Given the most general $T^{\prime}$
field content in~\eqnref{eqn:Lqrkyuk}, we allow a fully general VEV
structure,
\begin{equation}
\langle \alpha \rangle = \Lambda \left( \begin{array}{c} 
\alpha_{1} \\ \alpha_{2} \\ \alpha_{3} \\
\end{array} \right) 
\; , \quad
\langle \zeta \rangle = \Lambda \zeta_{0} 
\; ,  \quad
\langle \chi_U \rangle = \Lambda \left( \begin{array}{c}
\chi_{U1} \\ \chi_{U2} \\
\end{array} \right) 
\; , \quad
\langle \eta_U \rangle = \Lambda \left( \begin{array}{c}
\eta_{U1} \\ \eta_{U2} \\
\end{array} \right) \ ,
\end{equation}
which completely illustrates the maximum number of new parameters.
This allows us to write down the resulting 4D up-type mass
matrix\footnote{When performing the $2 \otimes 2 \otimes 3$
contraction in the upper $2 \times 2$ block of $M_U$, it may seem that
one could contract in two ways, $(2 \otimes 2) \otimes 3$ or $2
\otimes (2 \otimes 3)$, whereby the different Clebsh-Gordan
coefficients of the two contractions would lead to different
contributions.  This is not the case, however, since the first
contraction can be rescaled by $(1+i)$ to then become identical to the
second contraction.}
\begin{equation}
M_U = v \left( \begin{array}{ccc}
i \alpha_3 y_{12}^U f(c_{Q_{12}}, c_U) & 
\left[ \left( \frac{1 - i}{2} \right) \alpha_1 - \zeta_{0} \right] 
y_{12}^U f(c_{Q_{12}}, c_U) & \chi_{U2} y_3^U f(c_{Q_3}, c_U) \\
\left[ \left( \frac{1 - i}{2} \right) \alpha_1 + \zeta_{0} \right]
y_{12}^U f(c_{Q_{12}}, c_U) & \alpha_2 y_{12}^U f(c_{Q_{12}}, c_U) 
& -\chi_{U1} y_3^U f(c_{Q_3}, c_U) \\
\eta_{U2} y_{12}^T f(c_{Q_{12}}, c_T) & 
-\eta_{U1} y_{12}^T f(c_{Q_{12}}, c_T) & y_3^T f(c_{Q_3}, c_T) \\
\end{array} \right) \ .
\label{eqn:Mupgen}
\end{equation}

When performing the fit to the SM, we find that the general mass
matrix in~\eqnref{eqn:Mupgen} has an overabundance of parameters.  To
simplify the presentation, we assume the following, more restricted
VEV structure for the up-type and down-type flavons:
\begin{equation}
\langle \alpha \rangle = \langle \beta \rangle = \Lambda \alpha_{0} 
\left( \begin{array}{c} 
1 \\ 1 \\ 1 \\
\end{array} \right) 
\; , \quad
\langle \zeta \rangle = \langle \xi \rangle = \Lambda \zeta_{0}
\; ,  \quad
\langle \chi_r \rangle = \Lambda \chi_{0}^{r} \left( \begin{array}{c}
\cos \theta_r \\ \sin \theta_r \\
\end{array} \right) 
\; , \quad
\langle \eta_r \rangle = \Lambda \eta_{0}^{r} \left( \begin{array}{c}
1 \\ 0 \\
\end{array} \right) \ ,
\label{eqn:qrkvev}
\end{equation}
where $r = U$, $D$.  This VEV pattern leads to a 4D effective up-type
mass matrix
\begin{equation}
M_U = v \left( \begin{array}{ccc}
i \alpha_{0} f(c_{Q_{12}}, c_U) & 
\left[ \left( \frac{1-i}{2} \right) \alpha_{0} - \zeta_{0} \right] 
f(c_{Q_{12}}, c_U) 
& \chi_{0}^{U} \sin \theta_U f(c_{Q_3}, c_U) \\
\left[ \left( \frac{1-i}{2} \right) \alpha_{0} + \zeta_{0} \right] 
f(c_{Q_{12}}, c_U) 
& \alpha_{0} f(c_{Q_{12}}, c_U) & -\chi_{0}^{U} \cos \theta_U f(c_{Q_3}, c_U) \\
0 & -\eta_{0}^{U} f(c_{Q_{12}}, c_T) & y_3^T f(c_{Q_3}, c_T) \\
\end{array} \right)
\label{eqn:Umatrix}
\end{equation}
and a 4D effective down-type mass matrix
\begin{equation}
M_D = v \left( \begin{array}{ccc}
i \alpha_{0} f(c_{Q_{12}}, c_D) & 
\left[ \left( \frac{1-i}{2} \right) \alpha_{0} - \zeta_{0} \right] 
f(c_{Q_{12}}, c_D) 
& \chi_{0}^{D} \sin \theta_D f(c_{Q_3}, c_D) \\
\left[ \left( \frac{1-i}{2} \right) \alpha_{0} + \zeta_{0} \right] 
f(c_{Q_{12}}, c_D) 
& \alpha_{0} f(c_{Q_{12}}, c_D) & -\chi_{0}^{D} \cos \theta_D f(c_{Q_3}, c_D) \\
0 & -\eta_{0}^{D} f(c_{Q_{12}}, c_B) & y_3^B f(c_{Q_3}, c_B) \\
\end{array} \right) \ ,
\label{eqn:Dmatrix}
\end{equation}
where all other Yukawas have been set to 1.

\section{Results}
\label{sec:Results}
In this section, we present our numerical fits for the SM observed
values of fermion masses and mixings using the parameters of our
$T^{\prime}$ in RS model.  A naive counting for our model gives 8 (= 4
bulk + 3 Yukawa + 1 flavon) parameters for the charged leptons, 6[7]
(= 2 bulk + 2[3] Yukawa + 2 flavon) parameters for the Dirac [seesaw
realization] neutrinos, and 24 (= 6 bulk + 8 Yukawa + 10 flavon)
parameters for the quarks.  This is contrasted with the general
anarchy case, which has 45[39] parameters for the charged leptons and
Dirac [Majorana] neutrinos and 45 for the quarks.  In actuality, the
number of independent parameters for our model is much smaller: we
have 3 for the charged leptons, 2[2] for the neutrinos, and 11 for the
quarks, which compares to 36[30] for the anarchic leptons and 36 for
the anarchic quarks.

We briefly comment on the renormalization group (RG) effects to our
fit.  Our mass matrices for the SM fields are given at the (IR)
$T^\prime$ breaking scale, which we take to be $\sim$ 3 TeV.  For the
charged leptons and neutrinos, RG effects are negligible since the
running of Yukawa couplings from 3 TeV down to the $m_Z$ scale is
demonstrably small (cf. Table IV of Ref.~\cite{Fusaoka:1998vc}, where
charged lepton Yukawa coupling running from $10^9$ GeV to $m_Z$ is
less than a 10\% effect).  The mixing of neutrinos is also negligibly
affected by RG running since the neutrino masses at the $T^\prime$
scale are not sufficiently degenerate to enhance
mixing~\cite{Babu:1993qv}.  Quark masses, however, acquire
non-negligible corrections from running, while the corresponding
corrections to quark mixings are expected to be
small~\cite{Fusaoka:1998vc}.  Thus, we will fit to the charged leptons
masses at $m_Z$~\cite{Amsler:2008zzb}, the low energy neutrino mixing
data~\cite{Schwetz:2008er}, the quark masses at $ \sim 3$ TeV
(cf. Table 1 of Ref.~\cite{Csaki:2008zd}), and the CKM matrix at
$m_Z$~\cite{Charles:2004jd}.

We now fit the entire SM using our 16 independent parameters.  In the
charged lepton sector, all 5D Yukawas are set to 1, and we have $c_L =
0.40000$, $c_e = 0.82925$, $c_{\mu} = 0.66496$, $c_{\tau} = 0.57126$,
and $\phi_0 = 1$ as our input parameters.  These values give, at
$m_Z$, an electron mass of 511.1 keV, muon of 105.7 MeV, and a tau of
1.777 GeV, which are consistent with the experimental
values~\cite{Amsler:2008zzb} of $m_e = 510.998$ keV, $m_{\mu} =
105.658$ MeV, $m_{\tau} = 1.77684 \pm 0.00017$ GeV.

In the neutrino sector, for the Dirac case, we use the value of $c_L$
above, $c_N = 1.27000$, $\phi_{0, \text{ Dc}}^{\prime} = -0.1768$,
$\sigma_{0, \text{ Dc}} = 0.0944$, and we set both 5D Yukawas set to
1.  These parameters give absolute neutrino masses of $m_1 = -0.01563$
eV, $m_2 = 0.01791$ eV, and $m_3 = -0.05145$ eV. These correspond to
mass squared differences of $\Delta m_{21}^2 = 7.6370 \cdot 10^{-5}$
eV$^2$ and $\Delta m_{31}^2 = 2.4031 \cdot 10^{-3}$ eV$^2$, which are
in good agreement with the experimental results~\cite{Schwetz:2008er},
$\Delta m_{\text{sol}}^2 = 7.65_{-0.20}^{+0.23} \cdot 10^{-5}$ eV$^2$
for solar neutrino oscillation and $| \Delta m_{\text{atm}}^2 | =
2.40_{-0.11}^{+0.12} \cdot 10^{-3}$ eV$^2$ from atmospheric neutrinos.

In the seesaw realization of our model, to produce a normal hierarchy,
we use the value of $c_L$ above, $c_N = 0.40000$, $\phi_{0, \text{
SS}}^{\prime} = 0.07427$, $\sigma_{0, \text{ SS}} = 0.06191$, and we
set all three 5D Yukawas to 1.  This gives $m_1 = 0.004465$ eV, $m_2 =
0.009821$ eV, and $m_3 = 0.04919$ eV, and also we get $\Delta m_{21}^2
= 7.652 \cdot 10^{-5}$ eV$^2$ and $\Delta m_{31}^2 = 2.4001 \cdot
10^{-3}$ eV$^2$.  An inverted hierarchy solution arises if we use $c_N
= 0.40000$, $\phi_{0, \text{ SS}}^{\prime} = 0.02321$, $\sigma_{0,
\text{ SS}} = -0.0115241$, and again assign all Yukawas to be 1.  The
absolute masses are now $m_1 = 0.05203$ eV, $m_2 = -0.05276$ eV, and
$m_3 = 0.01751$, and the mass squared differences become $\Delta
m_{21}^2 = 7.656 \cdot 10^{-5}$ eV$^2$ and $\Delta m_{31}^2 = -2.4009
\cdot 10^{-3}$ eV$^2$.  Both seesaw solutions satisfy the current
experimental bounds quoted above.

For the quarks, we have the following input values for the flavon VEVs
and Yukawa couplings: $\alpha_{0} = -0.00143 + 0.00104i$, $\zeta_{0} =
0.00200$, $\chi_{0}^{U} = \eta_{0}^{U} = -0.448$, $\theta_U = 0.181
\pi$, $\chi_{0}^{D} = -0.00230$, $\theta_D = 0.1135 \pi$, $\eta_{0}^{D}
= -0.540 - 0.540i$, $y_3^T = 1.00$, and $y_3^B = 0.060$, with all
other Yukawa coupling constants set to 1.  In addition, the bulk mass
terms are $c_{Q_{12}} = 0.503$, $c_{Q3} = 0.150$, $c_U = 0.512$, $c_T
= -0.350$, $c_D = 0.503$, and $c_B = 0.508$.  These input parameters
give, at the (IR) $T^\prime$ breaking scale of 3 TeV, an up quark mass
of $1.49$ MeV, a charm mass of $0.541$ GeV, and a top mass of $134.8$
GeV.  The down-type quark masses are predicted to be $2.92$ MeV,
$36.6$ MeV, and $2.41$ GeV.  These masses are within the bounds of
Table 1 of Ref.~\cite{Csaki:2008zd}: $m_u = 0.75-1.5$ MeV, $m_c
= 0.56 \pm 0.04$ GeV, $m_t = 136.2 \pm 3.1$ GeV, $m_d = 2-4$
MeV, $m_s = 47 \pm 12$ MeV, $m_b = 2.4 \pm 0.04$ GeV.

The resulting CKM matrix from these input parameters is given by
\begin{equation}
V_{\text{CKM, th}} = \left( \begin{array}{ccc}
0.974282 e^{-0.0558i} & 0.225305 e^{-0.381i} & 0.003464 e^{1.31i} \\
0.225147 e^{-2.76i} & 0.973485 e^{0.0557i} & 0.040450 e^{3.13i} \\
0.00910164 e^{-3.12i} & 0.0395649 e^{0.0865i} & 0.999176 e^{0.0000095i} \\
\end{array} \right) \ .
\end{equation}
The absolute values of the CKM matrix elements agree with experimental
values at $m_Z$ within $3\sigma$~\cite{Charles:2004jd}:
\begin{equation}
\left| V_{\text{CKM, ex}} \right| = \left( \begin{array}{ccc}
0.97433_{-0.00052}^{+0.00052} & 0.2251_{-0.0022}^{+0.0022} & 
0.00351_{-0.00032}^{+0.00044} \\
0.2250_{-0.0022}^{+0.0022} & 0.97349_{-0.00052}^{+0.00053} & 
0.0412_{-0.0019}^{+0.0011} \\
0.00859_{-0.00064}^{+0.00057} & 0.0404_{-0.0020}^{+0.0011} & 
0.999146_{-0.000047}^{+0.000078} \\
\end{array} \right) \; .
\end{equation}
In addition, we have a predictions for $CP$ violation in the quark and
lepton sectors.  For the quark sector, our model predicts the
following value for the Jarlskog invariant,
\begin{equation}
J_{\text{th}} \equiv \text{Im } V_{ud} V_{cs} V_{us}^* V_{cd}^* = 3.02
\times 10^{-5} \ ,
\end{equation}
which is within the $3 \sigma$ uncertainty of the experimental
value~\cite{Charles:2004jd},
\begin{equation}
J_{\text{ex}} = 2.93_{-0.25}^{+0.45} \times 10^{-5} \ .
\end{equation}
We remark that, in our model, this value arises from a combination of
both complex Clebsch-Gordan coefficients~\cite{Chen:2009gf} and
complex VEVs of $T^{\prime}$ flavon fields (which are
indistinguishable from complex Yukawa coefficients).  For the leptons,
our Dirac mass matrices are completely real and diagonal, giving a
prediction of a vanishing leptonic Jarlskog.

In the absence of the FCNCs at tree-level at the renormalizable level
due to the $T^{\prime}$ family symmetry, the leading contributions to
flavor-violating $Z$ couplings are due to the dim-6 operators induced
by the mixing of fermion zero mode and its KK modes. For the first KK
mode, which gives the least suppressed contributions, these dim-6
operators lead to flavor violating $Z$ couplings, $Z \psi_{j}^{(0)}
\psi_{k}^{(0)}$.  Normalized to the SM $Z$ coupling, these higher
order effects contribute the following factor,
\begin{equation}
(\langle \alpha_0 \rangle + \langle \zeta_0 \rangle)^2
\frac{ (f_i^{(1)})^2 f_j^{(0)} f_k^{(0)} }{4 \pi^2 k^2 R^2}
\exp (2 \pi k R) \frac{ v^4 }{M_{KK}^2} \; ,
\end{equation}
where $f_{i}^{(1)}$ and $f_{j}^{(0)}$ are the wavefunction profiles of
the first KK mode and the zero mode of the fermion.  Numerically, for
the $u-c$ transition, the contribution is $2.965 \times 10^{-6}$ times
the regular $Z$ coupling.  The $d-s$ flavor violating transition
contributes $4.156 \times 10^{-6}$ times the regular $Z$ coupling,
assuming the first KK mass scale $\sim 3$ TeV. These higher order
effects are thus highly suppressed and are allowed by the experiments.

\section{Conclusion}
\label{sec:Conc}
We have proposed a Randall-Sundrum Model with a bulk $T^{\prime}$
family symmetry. The $T^{\prime}$ symmetry gives rise to a TBM mixing
matrix for the neutrinos and a realistic quark CKM matrix.  In the
lepton sector, exact neutrino tri-bimaximal mixing is generated due to
the group theoretical CG coefficients of $T^{\prime}$.  Since the
neutrino mass matrix is form diagonal, the neutrino mass eigenvalues
are decoupled from its mixing. This thus alleviates the tension
generally present in the anarchical scenarios between generating large
neutrino mixing angles and their hierarchical masses (the hierarchy
among the masses are determined by the flavon VEVs.)  For the charged
leptons, even though all three left-handed doublets have a common bulk
mass terms, the mass hierarchy among them are generated due to the
wave function profiles of the right-handed charged leptons.  In the
quark sector, the mass hierarchy between the first and second
generations are due to the structure of the $T^{\prime}$ flavon VEVs,
and the realistic CKM mixing arises due to both the flavon VEV pattern
and the wave function profiles.

We emphasize that the $T^{\prime}$ representation assignments required
for giving realistic masses and mixing patterns automatically forbid
all leptonic tree-level FCNCs and those involving the first and the
second generations of quarks, which are present in generic RS
models. As a result, a low scale for the first KK mass scale can be
allowed, rendering the RS model a viable solution to the gauge
hierarchy problem and making it testable at collider experiments.

\section*{Acknowledgements}

FY would like to thank the organizers and participants of the
Theoretical Advanced Study Institute (TASI 2009), hosted at the
University of Colorado at Boulder, where part of this work was
completed.  The work of M-CC was supported, in part, by the National
Science Foundation under grant No. PHY-0709742.  The work of KTM was
supported, in part, by the Department of Energy under grant
No. DE-FG02-04ER41290.

\begin{appendix}
\section{$T^{\prime}$ Family Symmetry}
\label{app:Tp}

The $T^{\prime}$ group is the double covering of the tetrahedral group
$A_4$, in an analogous way that $SU(2)$ is the double covering of
$SO(3)$.  It has 24 elements and two generators, $S$ and $T$.  It
contains three inequivalent, irreducible 1-dimensional
representations, three 2-dimensional representations, and one
3-dimensional representation.  The generators satisfy the following
algebra
\begin{equation}
S^2 = R, \qquad T^3 = 1, \qquad (ST)^3 = 1, \qquad R^2 = 1
\end{equation}
where $R = 1$ for the 1-dimensional and 3-dimensional representations,
and $R = -1$ for the two-dimensional representations. This can be
understood from the nomenclature that the 1-dimensional and
3-dimensional representations are vectorial representations, while the
2-dimensional representations are spinorial.  Just like spinors in
4-dimensions, the 2-dimensional representations acquire an extra $-1$
after rotation in $T^{\prime}$ space (or, analogously, $SU(2)$ space)
by $2 \pi$.  It is interesting to note that this feature generates
imaginary CG coefficients, which can be a source of CP
violation~\cite{Chen:2009gf}.  Using the conventions from
Ref.~\cite{Feruglio:2007uu}, the generators can be chosen as follows:
\begin{equation}
\begin{array}{cllll}
1   &\qquad& S = 1,   &\quad& T = 1, \\
1'  &\qquad& S = 1,   &\quad& T = \omega, \\
1'' &\qquad& S = 1,   &\quad& T = \omega^2, \\
2   &\qquad& S = A_1, &\quad& T = \omega A_2, \\
2'  &\qquad& S = A_1, &\quad& T = \omega^2 A_2, \\
2'' &\qquad& S = A_1, &\quad& T = A_2, \\
3   &\qquad& S = \frac{1}{3} \left( \begin{array}{ccc}
-1 & 2 \omega & 2 \omega^2 \\
2 \omega^2 & -1 & 2 \omega \\
2 \omega & 2 \omega^2 & -1 \\
\end{array} \right), &\quad&
T = \left( \begin{array}{ccc}
1 & 0 & 0 \\
0 & \omega & 0 \\
0 & 0 & \omega^2 \\
\end{array} \right) \,
\end{array}
\end{equation}
where the matrices $A_1$ and $A_2$ are,
\begin{equation}
A_1 = -\frac{1}{\sqrt{3}} \left( \begin{array}{cc}
i & \sqrt{2} e^{i \pi / 12} \\
-\sqrt{2} e^{-i \pi / 12} & i \\
\end{array} \right), \quad
A_2 = \left( \begin{array}{cc}
\omega & 0 \\
0 & 1 \\
\end{array} \right) \ .
\end{equation}

We briefly present the product rules relevant for our choice of
representation assignments in Table \ref{table:Tpassign} and the
Lagrangian specified in~\eqnref{eqn:Llep} and \eqnref{eqn:Lqrk}.  In
particular, the product of $3 \otimes 3$ appears in both the lepton
and quark sectors.  In the following, $\alpha_i$ denotes the $i$th
component of the first representation in the product, while $\beta_j$
denotes the $j$th component of the second representation in the
product.  We have
\begin{equation}
3 \otimes 3 = 3_S \oplus 3_A \oplus 1 \oplus 1' \oplus 1'' \ ,
\end{equation}
where
\begin{equation}
\begin{array}{l}
3_S = \frac{1}{3} \left( \begin{array}{c} 
2 \alpha_1 \beta_1 - \alpha_2 \beta_3 - \alpha_3 \beta_2 \\
2 \alpha_3 \beta_3 - \alpha_1 \beta_2 - \alpha_2 \beta_1 \\
2 \alpha_2 \beta_2 - \alpha_1 \beta_3 - \alpha_3 \beta_1 \\
\end{array} \right), \quad
3_A = \frac{1}{2} \left( \begin{array}{c}
\alpha_2 \beta_3 - \alpha_3 \beta_2 \\
\alpha_1 \beta_2 - \alpha_2 \beta_1 \\
\alpha_3 \beta_1 - \alpha_1 \beta_3 \\
\end{array} \right), \\
1 = \alpha_1 \beta_1 + \alpha_2 \beta_3 + \alpha_3 \beta_2, \\
1' = \alpha_3 \beta_3 + \alpha_1 \beta_2 + \alpha_2 \beta_1, \\
1'' = \alpha_2 \beta_2 + \alpha_1 \beta_3 + \alpha_3 \beta_1 \ . \\
\end{array}
\end{equation}
We remark that the factors of $\frac{1}{3}$ and $\frac{1}{2}$ in the
triplet representations of the direct sum are normalization
coefficients.

From the quark sector, we also require the products of $2 \otimes
2$ and $2 \otimes 3$.  The first product is
\begin{equation}
2 \otimes 2 = 3 \oplus 1
\end{equation}
where the corresponding CG coefficients are
\begin{equation}
3 = \left( \begin{array}{c}
\frac{1 - i}{2} \left( \alpha_1 \beta_2 + \alpha_2 \beta_1 \right) \\
i \alpha_1 \beta_1 \\
\alpha_2 \beta_2 \\
\end{array} \right), \qquad
1 = \alpha_1 \beta_2 - \alpha_2 \beta_1 \ ,
\end{equation}
while the second product is
\begin{equation}
2 \otimes 3 = 2 \oplus 2' \oplus 2''
\end{equation}
where
\begin{equation}
2 = \left( \begin{array}{c}
\left( 1 + i \right) \alpha_2 \beta_2 + \alpha_1 \beta_1 \\
\left( 1 - i \right) \alpha_1 \beta_3 - \alpha_2 \beta_1 \\
\end{array} \right) \ .
\end{equation}
We omit the other terms in the direct sums since they do not contract
to give pure singlets for the Lagrangian in~\eqnref{eqn:Llep} and
\eqnref{eqn:Lqrk}.  The remaining singlet contractions are
straightforward and detailed in~\cite{Feruglio:2007uu}.  From here,
some algebra on the Lagrangian gives the mass matrix structure
of~\eqnref{eqn:Lmatrix},~\eqnref{eqn:Nmatrix1},~\eqnref{eqn:Nmatrix2a},~\eqnref{eqn:Nmatrix2b},
and ~\eqnref{eqn:Mupgen}.

\end{appendix}

\end{document}